\documentclass[graybox]{svmult}

\usepackage{mathptmx}       
\usepackage{helvet}         
\usepackage{courier}        
\usepackage{type1cm}        
\usepackage{makeidx}         
\usepackage{graphicx}        
\usepackage{multicol}        
\usepackage[bottom]{footmisc}

\usepackage{amssymb}
\usepackage{amsmath}
\usepackage{bbm}

\newcommand{\ii}{\mathrm{i}}

\newcommand{\de}{\partial}
\newcommand{\alp}{\alpha^{(1,1)}}

\makeindex             

\begin{document}

\title*{Effective non-linear dynamics of binary condensates and open problems}
\authorrunning{Alessandro Olgiati}
\author{Alessandro Olgiati,\\$\,$ \footnotesize\\SISSA - International School for Advanced Studies,\\ via Bonomea 265, Trieste, Italy\\\email{aolgiati@sissa.it}}
\maketitle

\abstract*{We report on a recent result concerning the effective dynamics for a mixture of Bose-Einstein condensates, a class of systems much studied in physics and receiving a large amount of attention in the recent literature in mathematical physics; for such models, the effective dynamics is described by a coupled system of non-linear Sch\"odinger equations. After reviewing and commenting our proof in the mean-field regime from a previous paper, we collect the main details needed to obtain the rigorous derivation of the effective dynamics in the Gross-Pitaevskii scaling limit.}

\abstract{We report on a recent result concerning the effective dynamics for a mixture of Bose-Einstein condensates, a class of systems much studied in physics and receiving a large amount of attention in the recent literature in mathematical physics; for such models, the effective dynamics is described by a coupled system of non-linear Sch\"odinger equations. After reviewing and commenting our proof in the mean-field regime from a previous paper, we collect the main details needed to obtain the rigorous derivation of the effective dynamics in the Gross-Pitaevskii scaling limit.}

\section{Introduction}
Bose-Einstein condensation is the physical phenomenon according to which a macroscopic number of bosons collapse onto the same quantum state. This was first predicted theoretically in the 1920's and then widely studied both in physics and mathematics in the later decades; the topic received a further strong boost since the mid 1990's, when the first condensates were produced in experiments.

Mathematically, to a system of $N$ identical bosons is associated the Hilbert space $L^2(\mathbb{R}^3)^{\otimes_\text{sym}N}$ and states are positive trace-class operators $\gamma_N$ on such space, with unit trace. The notion of condensation is appropriately described in terms of the corresponding one-body reduced density matrix, or one-body marginal,
\begin{equation} \label{eq:reduced_density_matrix}
\gamma^{(1)}_N=\text{Tr}_{N-1}\gamma_N,
\end{equation}
where the degrees of freedom 2 to $N$ are traced out; the operation $\text{Tr}_{N-1}$ in \eqref{eq:reduced_density_matrix} is called the partial trace. Thus, given a $N$-body density matrix $\gamma_N$ of the system, and a pure state $u\in L^2(\mathbb{R}^3)$, one says that $\gamma_N$ exhibits complete asymptotic condensation on the condensate wave-function $u$ if
\begin{equation} \label{eq:condensation}
\lim_{N\rightarrow\infty}\gamma_N^{(1)}=\;|u\rangle\langle u|.
\end{equation}
Since the limit in \eqref{eq:condensation} is a rank-one projection, weak convergence implies trace-norm convergence, and thus, the limit can be considered in any of such topologies.

Within this framework, a problem naturally arising is the proof of persistence of condensation under the dynamics generated by some many-body Hamiltonian. Thus, given a time-evolution governed by $H_N$, and the flow
\[
\gamma_N\mapsto \gamma_{N,t}=e^{-\ii t H_N}\gamma_N e^{\ii t H_N},
\]
one would like to prove that
\begin{equation} \label{eq:propagation_of_chaos}
\gamma_{N}^{(1)}\;\simeq\;|u_0\rangle\langle u_0|\quad\Rightarrow\quad \gamma_{N,t}^{(1)}\simeq\;|u_t\rangle\langle u_t|.
\end{equation}
The interest in a result like \eqref{eq:propagation_of_chaos} is manifest: a large system is well approximated by a single-particle orbital, an object much more manageable in computations and informative when one-body observables are considered. The price to pay is that in the limit the interparticle interactions result in a non-linearity, or self-interaction term; hence, a typical equation for $u_t$ is
\[
\ii \de_tu_t= -\Delta u_t+\mathcal{N}(u_t)u_t,
\] 
where, as said, $\mathcal{N}(.)$ accounts for the effective two-body potential via a cubic self-interaction. We refer to the review~\cite{Benedikter-Porta-Schlein-2015} for a comprehensive outlook on the problem. It has to be remarked that this class of problems has involved many different techniques, with tools from operator theory, measure theory and kinetic theory.

\section{Two-component condensates}\label{sect:multi}
A consistent part of both theoretical and experimental studies on Bose-Einstein condensation is devoted to systems in which two (or more) components interact; such systems are usually referred to as two-component condensates (respectively multi-component condensates). This can be attained in multiple ways: either by considering bosons occupying different hyperfine spin states~\cite{MBGCW-1997,Stamper-Kurn_Ketterke_et_al_PRL-1998} (spinor condensates) or by considering different atomic species~\cite{Modugno-PRL-2002} (mixture condensates); in the case of different spin states, one can also account for transitions between the two components, for example by turning on an external magnetic field or a spin-spin interaction (this is discussed in Sect. \ref{sect:open}). Physical evidence suggests that the dynamics of a multi-component condensate is governed by a coupled system of non-linear Schr\"odinger equations (see~\cite[Section 21]{ps2016}), the unknowns being the condensate wave-functions of each component.

In this work we consider the case of the mixture condensate, namely a system consisting of $N_1$ identical bosons of some atomic species $A$ and $N_2$ identical bosons of some (different) species $B$; the Hilbert space of such system is
\begin{equation} \label{eq:hilbert_space}
\mathcal{H}_{N_1,N_2}=L^2_{\text{sym}}(\mathbb{R}^{3N_1},dx_1\dots dx_{N_1})\otimes L^2_{\text{sym}}(\mathbb{R}^{3N_1},dy_1\dots dy_{N_2}).
\end{equation}
We want to consider states of such system in which condensation is present in each component: this can be monitored by means of a \emph{``double'' reduced density matrix}. For each state $\gamma_{N_1,N_2}$ of the system, we define the trace-class operator
\begin{equation} \label{eq:def_double_partial_trace-KERNEL}
\gamma^{(1,1)}_{N_1,N_2}=\text{Tr}_{N_1-1}\otimes\text{Tr}_{N_2-1}\gamma_{N_1,N_2},
\end{equation}
acting on the space $L^2(\mathbb{R}^3,dx)\otimes L^2(\mathbb{R}^3,dy)$ of one particle of type A and one of type B.

In this setting, one can extend the notion of condensation, namely, one says $\gamma_{N_1,N_2}$ exhibits complete condensation in both components, with condensate functions $u$ and $v$, if
\begin{equation}
\lim_{\substack{N_1\to\infty \\ N_2\to\infty}}\gamma_{N_1,N_2}^{(1,1)}\;=\;|u\otimes v\rangle\langle u\otimes v|\;=\;|u\rangle\langle u|\otimes |v\rangle\langle v|. \label{eq:def_100BEC-2component}
\end{equation}
In analogy to the one-component case, it is of interest to investigate the persistence of condensation simultaneously in each component. Of course one has to specify a Hamiltonian generating the time-evolution; moreover, since at the moment no result is attainable in a genuine thermodynamic limit of large system, the Hamiltonian must be chosen together with a scaling prescription that mimics the true limit.

\subsection{Mean-field regime}\label{sect:mf}

For the multi-component system built in Section \ref{sect:multi}, we define the three-dimensional mean-field Hamiltonian
\begin{equation}\label{eq:HN1N2}
\begin{split}
H_{N_1,N_2}\;=\;&\sum_{i=1}^{N_1}(-\Delta_{x_i})+\frac{1}{N_1}\sum_{i<j}^{N_1}V_1(x_i-x_j) \\
+\;&\sum_{r=1}^{N_2}(-\Delta_{y_r})+\frac{1}{N_2}\sum_{r<s}^{N_2}V_2(y_r-y_s) \\
&\quad +\frac{1}{N_1+N_2}\sum_{i=1}^{N_1}\sum_{r=1}^{N_2}V_{12}(x_i-y_r),
\end{split}
\end{equation}
where the variables $x_1,\dots x_{N_1},y_1\dots y_{N_2}$ are referred to the ones in Eq. \eqref{eq:hilbert_space}.

Throughout this paper, we will consider the case in which $N_1$ and $N_2$ scale in such a way that their ratio is asymptotically constant, namely there exist constants $c_1,c_2>0$ such that
\begin{equation}\label{eq:def_c12}
c_i=\lim_{\substack{N_1\rightarrow\infty\\ N_2\rightarrow\infty}}\frac{N_i}{N_1+N_2},\quad i=1,2.
\end{equation}
For simplicity of presentation, we assume that \eqref{eq:def_c12} holds identically for every fixed $N_1$ and $N_2$, and not only in the limit; this stronger assumption could easily be removed.
Under such assumptions, it is easy to see that our choice of the mean-field pre-factors $N_1^{-1}$, $N_2^{-1}$, $(N_1+N_2)^{-1}$ ensures all terms in \eqref{eq:HN1N2} to remain of the same order $O(N_1+N_2)$. Of course, one could argue that many other choices would ensure this behavior, for example a common $(N_1N_2)^{-1/2}$ factor; the reader can refer to Sect. 4 in ~\cite{mo-mf2016} for a discussion of why the choice in \eqref{eq:HN1N2} is \emph{the} physically relevant mean-field scaling.

Our result is the proof of persistence of condensation under the dynamics generated by \eqref{eq:HN1N2}, namely
\begin{equation} \label{eq:result}
\gamma_{N_1,N_2}^{(1,1)}(0)\;\simeq\;|u_0\otimes v_0\rangle\langle u_0\otimes v_0|\quad\Rightarrow\quad \gamma_{N_1,N_2}^{(1,1)}(t)\simeq\;|u_t\otimes v_t\rangle\langle u_t\otimes v_t|,
\end{equation}
where $(u_t,v_t)$ solves the initial value problem
\begin{equation}\label{eq:Hartree_system}
\begin{split}
\ii\partial_t u_t\;&=\;-\Delta u_t + (V_1*|u_t|^2) u_t + c_2 (V_{12}*|v_t|^2) u_t \\
\ii\partial_t v_t\;&=\;-\Delta v_t + (V_2*|v_t|^2) v_t + c_1 (V_{12}*|u_t|^2) v_t,
\end{split}
\end{equation}
with initial datum $(u_0,v_0)$. 

Let us now state the assumptions on $V_j$ and $(u_0,v_0)$ under which it is possible to prove \eqref{eq:result}.

\begin{itemize}
	 \item{(A1)} The potentials $V_j$, $j\in\{1,2,12\}$ are real-valued, even, and such that
	 \begin{equation}\label{eq:assumptions_on_V_alternative}
	 \begin{split}
	 \|V_j*|\phi|^2\|_\infty\;&\lesssim\;\|\phi\|^2_{H^1}\qquad\forall\phi\in H^1(\mathbb{R}^3)\qquad j=1,2,12 \\
	 \|V_j^2*|\phi|^2\|_\infty\;&\lesssim\;\|\phi\|^2_{H^1}\qquad\forall\phi\in H^1(\mathbb{R}^3)\qquad j=1,2,12 .
	 \end{split}
	 \end{equation}
	 \item{(A2)} The initial data for the system \eqref{eq:Hartree_system} are $u(0)=u_0$ and $v(0)=v_0$ for given functions $u_0,v_0\in H^1(\mathbb{R}^3)$ with $\|u_0\|_2=\|v_0\|_2=1$. By general theory, this is enough to have a unique global-in-time solution
	 \begin{equation}\label{eq:solutions}
	 (u_t,v_t)\;\in\;C(\mathbb{R},H^1(\mathbb{R}^3)\oplus H^1(\mathbb{R}^3))\cap C^1(\mathbb{R},H^{-1}(\mathbb{R}^3)\oplus H^{-1}(\mathbb{R}^3)).
	 \end{equation}
	 \item{(A3)} The many-body initial datum is $\Psi_{N_1,N_2}\in\mathcal{D}[H_{N_1,N_2}]\cap \mathcal{H}_{N_1,N_2,\mathrm{sym}}$ with $\|\Psi_{N_1,N_2}\|_2=1$.
\end{itemize}

Let $\Psi_{N_1,N_2}(t):=e^{-\ii t H_{N_1,N_2}}\Psi_{N_1,N_2}$ be the unique solution in $C(\mathbb{R},\mathcal{D}[H_{N_1,N_2}]\cap \mathcal{H}_{N_1,N_2,\mathrm{sym}})$ to the many-body Schr\"{o}dinger equation
\begin{equation}\label{eq:manybody-Schr}
\ii\partial_t\Psi_{N_1,N_2}(t)\;=\;H_{N_1,N_2}\Psi_{N_1,N_2}(t)\,,\qquad \Psi_{N_1,N_2}(0)=\Psi_{N_1,N_2}\,,
\end{equation}
and let $(u_t,v_t)$ be the unique solution to the system of coupled NLS \eqref{eq:Hartree_system} as in \eqref{eq:solutions}.
Our main result in the mean-field regime is the following Theorem.

\begin{theorem}[\cite{mo-mf2016}] \label{theorem:main}
	Consider a two-species bosonic system under assumptions (A1)-(A3) above. Let  $\gamma^{(1,1)}_{N_1,N_2}(t)$ be the double reduced density matrix associated with $\Psi_{N_1,N_2}(t)$, given by \eqref{eq:def_double_partial_trace-KERNEL}, and define
	\begin{equation}\label{eq:def_alpha}
	\alpha^{(1,1)}_{N_1,N_2}(t)\;:=\;1-\big\langle u_t\otimes v_t\:,\:\gamma_{N_1,N_2}^{(1,1)}(t)\;u_t\otimes v_t\big\rangle\,.
	\end{equation}
	Then
	\begin{equation}\label{tesi}
	\alpha^{(1,1)}_{N_1,N_2}(t)\;\leqslant\;\left(\alpha_{N_1,N_2}^{(1,1)}(0)+\frac{1}{N_1+N_2}\right)e^{\,f(t)},
	\end{equation}
	where $f$ does not depend on $N$.
\end{theorem}

\begin{corollary}[\cite{mo-mf2016}]\label{corollary:main}
	In the same hypothesis of Theorem \ref{theorem:main}, if
	\[
	\lim_{\substack{N_1\rightarrow\infty\\N_2\rightarrow\infty}}\gamma_{N_1,N_2}^{(1,1)}(0)=\;|u_0\otimes v_0\rangle\langle u_0\otimes v_0|,
	\]
	in trace norm, then
	\[
	\lim_{\substack{N_1\rightarrow\infty\\N_2\rightarrow\infty}}\gamma_{N_1,N_2}^{(1,1)}(t)=\;|u_t\otimes v_t\rangle\langle u_t\otimes v_t|,
	\]
	again in trace norm.
\end{corollary}

We show here the immediate proof of Corollary \ref{corollary:main}, postponing to Sect. \ref{section:proofmain} a sketch of the proof of Theorem \ref{theorem:main}, 

\begin{proof}[Corollary \ref{corollary:main}]
	The thesis follows from \eqref{tesi} using the chain of inequalities (see~\cite{mo-mf2016} eq. 3.7)
	\begin{equation} \label{eq:chain}
	\alpha_{N_1,N_2}^{(1,1)}(t)\leqslant \text{Tr}\Big|\gamma_{N_1,N_2}^{(1,1)}(t)-|u_t\otimes v_t\rangle\langle u_t\otimes v_t|\Big|\leqslant C \sqrt{ \alpha_{N_1,N_2}^{(1,1)}(t)}.
	\end{equation}
\end{proof}

A few remarks on the results we stated are in order.

\begin{remark}
Assumption (A1) covers, by Hardy inequality, the physically relevant case of Coulomb singularities $|x|^{-1}$.
\end{remark}

\begin{remark}
	To keep the exposition short and self-contained, we limited the class of Hamiltonians for which a result like Theorem \ref{theorem:main} holds; in particular, one could deal with several meaningful generalizations of the one-body operator $-\Delta$, as for example the magnetic Laplacian with external potential $-\Delta_A+U(x)$, or its semi-relativistic counterpart $(1-\Delta_A)^{1/2}+U(x)$, where $\Delta_A:=(\nabla-iA)^2$.
\end{remark}

\begin{remark}
	The second bound in \eqref{eq:chain} is not sharp: indeed, one could adapt a recent result~\cite{mpp-bog2016} and obtain convergence in trace norm \emph{with the same rate as the convergence of $\alpha_{N_1,N_2}^{(1,1)}$}. This, by \eqref{tesi}, implies that the total rate is the worst among the rates of $\alpha_{N_1,N_2}^{(1,1)}(0)$ and of $(N_1+N_2)^{-1}$.
\end{remark}

The functional $\alpha_{N_1,N_2}^{(1,1)}(t)$ is a two-component generalization of the one-component functional
\[
\alpha_N(t):=1-\langle\psi_N(t),p_1(t)\psi_N(t)\rangle,
\]
where
\begin{equation}\label{eq:def_p}
p_1(t):=|u_t\rangle_1\langle u_t|_1
\end{equation}
 is the projection onto the condensate wave-function in the variable $x_1$; for later convenience we also define the orthogonal complement to $p$ as
 \begin{equation}\label{eq:def_q}
 q_1(t):=\mathbbm{1}-p_1(t).
 \end{equation}
 Such a construction is the starting point of the so-called ``counting'' method introduced by Pickl in~\cite{p-simple2011} and by Knowles and Pickl in~\cite{kp-singular2010}. In those works, $\alpha_N(t)$ is used to prove trace-norm convergence with a quantitative rate for a wide class of potentials in the single component case.
 
 The meaning of Equation \eqref{eq:chain} (and of its one-component counterpart, see Lemma 2.3 in \cite{kp-singular2010}) is that $\alpha_{N_1,N_2}^{(1,1)}$ is a convenient indicator of condensation, namely its convergence to zero is tantamount as the convergence in trace norm to the condensate wave-function. In our two-component case, one could also argue that condensation can also be expressed in terms of one-component reduced density matrices, which can be defined as
 \begin{equation}
 \gamma_{N_1,N_2}^{(1,0)}=\text{Tr}_{N_1-1}\otimes\text{Tr}_{N_2}\gamma_{N_1,N_2},\qquad  \gamma_{N_1,N_2}^{(0,1)}=\text{Tr}_{N_1}\otimes\text{Tr}_{N_2-1}\gamma_{N_1,N_2}.
 \end{equation}
 The control of condensation by means of both $\gamma_{N_1,N_2}^{(1,0)}$ and $\gamma_{N_1,N_2}^{(0,1)}$ has been addressed by Heil \cite{h-thesis} (we also refer to \cite{ahh-hartree2017} for a more recent work); in Lemma 3.1 in \cite{mo-mf2016} we establish the bound
 \begin{equation}\label{eq:bound_g11_g10_g01}
 \begin{split}
 \max\big\{1-\langle u,&\gamma_{N_1,N_2}^{(1,0)} u\rangle\,,\,1-\langle v,\gamma_{N_1,N_2}^{(0,1)} v\rangle\big\}\;\leqslant\;1-\langle u\otimes v,\gamma_{N_1,N_2}^{(1,1)} u\otimes v\rangle \\
 &\leqslant\;(1-\langle u,\gamma_{N_1,N_2}^{(1,0)} u\rangle) + (1-\langle v,\gamma_{N_1,N_2}^{(0,1)} v\rangle),
 \end{split}
 \end{equation}
 which shows that our collective indicator $\gamma_{N_1,N_2}^{(1,1)}$ covers (and is in fact equivalent to) such a control.

\subsection{Gross-Pitaevskii regime} \label{sect:gp}
The mean-field result stated above can be extended to the more interesting and realistic Gross-Pitaevskii regime we describe in the following; in its essence, what we report already stems from the work \cite{mo-mf2016}. Nonetheless, we state here the result and present the main steps of the proof, in order to provide an explicit reference.

Consider the two-component Hamiltonian
\begin{equation}\label{eq:HN1N2GP}
\begin{split}
H_{N_1,N_2}\;=\;&\sum_{i=1}^{N_1}(-\Delta_{x_i})+N_1^{2}\sum_{i<j}^{N_1}V_1(N_1(x_i-x_j)) \\
+\;&\sum_{r=1}^{N_2}(-\Delta_{y_r})+N_2^{2}\sum_{r<s}^{N_2}V_2(N_2(y_r-y_s)) \\
&\quad +(N_1+N_2)^{2}\sum_{i=1}^{N_1}\sum_{r=1}^{N_2}V_{12}((N_1+N_2)(x_i-y_r)),
\end{split}
\end{equation}
where now the potentials are rescaled according to the Gross-Pitaevskii scaling. This implies very strong ($\sim N^{2}$) but rare interactions, since particles interact only when their distances are of order $ N^{-1}$, and this makes the regime quite different from the mean field in Sect. \ref{sect:mf}: whereas in mean field each particle only feels the average density of the whole gas, in the Gross-Pitaevskii regime interactions are very strong and effective only on short spatial scales. For this reason, this scaling is a much more realistic approximation for a gas in a zero temperature and high dilution regime.

One can prove a statement similar to Theorem \ref{theorem:main} also in this case, but with an amount of modifications. Indeed, now the limit \eqref{eq:result} holds for $(u_t,v_t)$ solutions to the \emph{local} system of NLS
\begin{equation}\label{eq:GP_system}
\begin{split}
\ii\partial_t u_t\;&=\;-\Delta u_t + 8 \pi a_1 |u_t|^2 u_t + c_2\, 8 \pi a_{12}|v_t|^2 u_t \\
\ii\partial_t v_t\;&=\;-\Delta v_t + 8\pi a_2|v_t|^2 v_t + c_1 \,8 \pi a_{12}|u_t|^2 v_t,
\end{split}
\end{equation}
where, for $j\in\{1,2,12\}$, $a_j$ is the $s$-wave scattering length of $V_j$.

Since, to treat the Gross-Pitaevskii case, one also has to take into account energy comparisons between many-body and effective dynamics, we define the following two functionals: the many-body energy functional
\begin{equation}
\mathcal{E}_{N_1,N_2}(\Psi_{N_1,N_2}):=\frac{1}{N_1+N_2}\langle\Psi_{N_1,N_2},H_{N_1,N_2}\Psi_{N_1,N_2}\rangle,
\end{equation}
and the Gross-Pitaevskii energy
\begin{equation}
\begin{split}
\mathcal{E}^{GP}(u,v):=&\langle u, -\Delta u\rangle+\langle v, -\Delta v\rangle+4\pi a_1 \langle u, |u|^2 u\rangle\\
&+4\pi a_2 \langle v, |v|^2 v\rangle+8\pi a_{12} \langle u, |v|^2 u \rangle.
\end{split}
\end{equation}

We suppose the following on the potential and on the initial data.

\begin{itemize}
	\item{(B1)} The potentials $V_\alpha$, $\alpha\in\{1,2,12\}$ are positive, spherically symmetric, compactly supported, $L^\infty$-functions.
	\item{(B2)} The initial data for the system \eqref{eq:Hartree_system} are $u(0)=u_0$ and $v(0)=v_0$ for given functions $u_0,v_0\in L^2(\mathbb{R}^3)$ with $\|u_0\|_2=\|v_0\|_2=1$ chosen such that the solution belongs to
	\[
	L^\infty\Big(\mathbb{R},H^2(\mathbb{R}^3)\oplus H^2(\mathbb{R}^3)\Big).
	\]
	\item{(B3)} The many-body initial datum is $\Psi_{N_1,N_2}\in\mathcal{D}[H_{N_1,N_2}]\cap \mathcal{H}_{N_1,N_2,\mathrm{sym}}$ with $\|\Psi_{N_1,N_2}\|_2=1$ and
	\[
	\lim_{\substack{N_1\rightarrow\infty\\N_2\rightarrow\infty}}\gamma_{N_1,N_2}^{(1,1)}=\;|u_0\otimes v_0\rangle\langle u_0\otimes v_0|.
	\]
	\item{(B4)} The sequence $\Psi_{N_1,N_2}$ satisfies
	\[
	\lim_{\substack{N_1\rightarrow\infty\\N_2\rightarrow\infty}}\mathcal{E}_{N_1,N_2}(\Psi_{N_1,N_2})=\mathcal{E}^{GP}(u_0,v_0).
	\]
\end{itemize}
Here is our main result.

\begin{theorem} \label{theorem:mainGP}
	Consider a two-species bosonic system under assumptions (B1)-(B4) above. Let  $\gamma^{(1,1)}_{N_1,N_2}(t)$ be the double reduced density matrix associated with $\Psi_{N_1,N_2}(t)$, given by \eqref{eq:def_double_partial_trace-KERNEL}. Then
	\begin{equation}\label{eq:tesiGP}
    \lim_{\substack{N_1\rightarrow\infty\\N_2\rightarrow\infty}}\gamma_{N_1,N_2}^{(1,1)}(t)=\;|u_t\otimes v_t\rangle\langle u_t\otimes v_t|,
	\end{equation}
	and
	\begin{equation}
	\lim_{\substack{N_1\rightarrow\infty\\N_2\rightarrow\infty}}\mathcal{E}_{N_1,N_2}(\Psi_{N_1,N_2}(t))=\mathcal{E}^{GP}(u_t,v_t),
	\end{equation}
	where $(u_t,v_t)$ are solutions of \eqref{eq:GP_system} with initial data $(u_0,v_0)$.
\end{theorem}
\begin{remark}
	A generalization of the technique used in the proof allows one to cover also the case of one-body Hamiltonians more general than $-\Delta$. This has been pointed out in the single component case in Remark 2.1 in \cite{p-external2015}; we refer the reader to \cite{o-magnetic} for a more detailed analysis of what is needed in order to adapt the argument to the relevant case of the magnetic Laplacian $\Delta_A=(\nabla-iA)^2$
\end{remark}
\begin{remark}
	Assumption (B1) on the potential is crucial in this formalism; with different techniques (see \cite{bdos-quant2012}) it is possible to consider potentials with some singularity and unbounded support. Conversely, the removal of the positivity condition is an important open problem in the subject; in \cite{p-positivity2010}, it is proven positivity can be removed for a much softer scaling than the one in \eqref{eq:HN1N2GP}.
\end{remark}

The one-component problem, namely the derivation of the Gross-Pitaevskii equation
\[
\ii\partial_t u_t\;=\;-\Delta u_t + 8 \pi a |u_t|^2 u_t,
\]
has been an important open problem in mathematical physics in recent years. It was first solved by Erd\H{o}s, Schlein and Yau in 2006 (see \cite{esy-2007} and \cite{esy-2010}); their proof was based on the BBGKY formalism and did not provide a convergence rate. Later results by Benedikter, de Oliveira and Schlein \cite{bdos-quant2012} and by Pickl \cite{p-external2015} relied on different techniques and allowed to get a quantitative control of the convergence.

\section{Proof of Theorem \ref{theorem:main}} \label{section:proofmain}

The strategy to get \eqref{tesi} is to establish an estimate of type
\begin{equation}\label{eq:gronwall}
\de_t{\alpha}_{N_1,N_2}^{(1,1)}(t)\leqslant\;f(t)\left(\alpha_{N_1,N_2}^{(1,1)}(t)+\frac{1}{N_1+N_2}\right),
\end{equation}
and then to apply Gr\"onwall lemma to get the result. The function $f$ will depend on the population ratios $c_1,c_2$ and on certain norms of the potentials $V_1,V_2,V_{12}$ and of the solutions $u_t$, $v_t$. For brevity, we will use from now on the shorthand notation $\alpha^{(1,1)}:={\alpha}_{N_1,N_2}^{(1,1)}(t)$.

One can show that our hypothesis certainly assure $\alp$ to be differentiable in time; its derivative can be shown to split into three pieces, each one of them containing only one potential, according to
\begin{equation}\label{eq:alpha-C-splitting}
\dot{\alpha}^{(1,1)}\;=\;\ii\,(C_{V_1}+C_{V_2}+C_{V_{12}}),
\end{equation}
with
\begin{equation}\label{eq:C-V1}
C_{V_1}:=\Big\langle\Psi,\Big[\Big(\dfrac{1}{N_1}\sum_{i<j}^{N_1}V_1(x_i-x_j)-\sum_{i=1}^{N_1}(V_1^u)_i\Big)^A ,\sum_{k=1}^{N_1}\sum_{\ell=1}^{N_2}\dfrac{\mathbbm{1}-p_k^A\,p_\ell^B}{N_1N_2}\Big]\Psi\Big\rangle,
\end{equation}
\begin{equation}\label{eq:C-V2}
C_{V_2}:=\Big\langle\Psi,\Big[\Big(\dfrac{1}{N_2}\sum_{r<s}^{N_2}V_2(y_r-y_s)-\sum_{r=1}^{N_2}(V_2^v)_r\Big)^B ,\sum_{k=1}^{N_1}\sum_{\ell=1}^{N_2}\dfrac{\mathbbm{1}-p_k^A\,p_\ell^B}{N_1N_2}\Big]\Psi\Big\rangle,
\end{equation}
\begin{equation}\label{eq:C-V12}
\begin{split}
C_{V_{12}}&=\Big\langle\Psi,\Big[\dfrac{1}{N_1+N_2}\sum_{i=1}^{N_1}\sum_{r=1}^{N_2}V_{12}(x_i-y_r)-c_2\sum_{i=1}^{N_1}(V_{12}^v)_i^A \\
&\qquad\qquad -c_1\sum_{r=1}^{N_2}(V_{12}^u)_r^B ,\sum_{k=1}^{N_1}\sum_{\ell=1}^{N_2}\dfrac{\mathbbm{1}-p_k^A\,p_\ell^B}{N_1N_2}\Big]\Psi\Big\rangle\,.
\end{split}
\end{equation}
Here and in what follows, the superscript $A$ (respectively $B$) indicates that $p_1^A$ acts on the first variable of the sector $A$, namely $x_1$ (respectively $y_1$).
Each of these three summands will be estimated in terms of $\alp$ and of $(N_1+N_2)^{-1}$ so as to obtain \eqref{eq:gronwall}. The terms $C_{V_1}$ and $C_{V_2}$ contain only infra-species interactions, and, for this reason, their estimate is less involved; the detailed proof can be found in ~\cite{mo-mf2016} (see also~\cite{kp-singular2010} for the single-component case).

To estimate $C_{V_{12}}$ one can exploit the bosonic symmetry of $\Psi$ and the definition of $c_j$ to obtain the bound
\begin{equation}\label{eq:CV12-simplified}
|C_{V_{12}}|\;\leqslant\;\frac{N_1N_2}{N_1+N_2}\,\Big|\Big\langle\Psi,\Big[(V_{12})_{11}-(V_{12}^v)_1^A-(V_{12}^u)_1^B ,\sum_{k=1}^{N_1}\sum_{\ell=1}^{N_2}\dfrac{p_k^A\,p_\ell^B}{N_1N_2}\Big]\Psi\Big\rangle\Big|\,. 
\end{equation}
At this point, one is free to insert, on both sides of the commutator, the identity
\begin{equation}\label{eq:add_idendity-AB}
\mathbbm{1}\;=\;(p_1^A+q_1^A)(p_1^B+q_1^B),
\end{equation}
with $p$ and $q$ as in \eqref{eq:def_p}, \eqref{eq:def_q}. The insertion clearly produces 16 terms, that we can split into two groups with a self-explanatory notation
\begin{equation}\label{eq:Lambda}
\begin{split}
\Lambda\;&:=\;(pp,pp)+[(pq,pq)+(qp,qp)]+(qq,qq) \\
&\qquad +\left[(pq,qp)+\text{complex conjugate}\,\right]
\end{split}
\end{equation}
and
\begin{equation}\label{eq:Omega}
\begin{split}
\Omega\;&:=\;(pp,qp)+(qp,qq)+(pp,qq)+(pp,pq)+(pq,qq) \\
&\qquad +\text{ complex conjugate }.
\end{split}
\end{equation}
The terms $(pp,pp)$, $(qq,qq)$, $(pq,pq)+(qp,qp)$ in \eqref{eq:Lambda} vanish identically, which can be easily checked using the fact that $p_1^Aq_1^A=0$; all the others could, in principle, provide some contribution to \eqref{eq:gronwall}. While we refer the reader to Sect. 5 in \cite{mo-mf2016} for the detailed computation, we try to sketch here how each term can be handled.

Since we need to reconstruct $\alp=1-\langle\Psi,p_1^Ap_1^B\Psi\rangle$ (as in \eqref{eq:gronwall}), we can make a clever use of every $q_1^A\Psi$ or $q_1^B\Psi$ in the non-vanishing terms: indeed, $\|q_1^A\Psi\|^2\leqslant\alp$. For this reason, when at least one $q$ from \eqref{eq:add_idendity-AB} appears on each side of the commutator, one only has to control the operator norm of $p_1V_{12}(x_1-y_1)$ and this allows to obtain the bound
\begin{equation}\label{eq:estimateII}
\begin{split}
\Big|\left[(pq,qp)+\text{c.c.}\,\right]&+\left[(qp,qq)+(pq,qq)+\text{c.c}\,\right]\Big|\\
&\leqslant \;f(t)\left(\alpha_{N_1,N_2}^{(1,1)}(t)+\frac{1}{N_1+N_2}\right).
\end{split}
\end{equation}
The term $(pp,qq)$ has the correct number of $q$'s too, but they appear on the same side, and this would not allow to extract $\|q_1^A\Psi\|^2$; however, one $q$ can be brought to the other side at the expense of some $(N_1+N_2)^{-1}$ smallness. This allows to obtain
\begin{equation}\label{eq:estimateIII}
\Big|(pp,qq)+\text{c.c.}\Big|\leqslant \;f(t)\left(\alpha_{N_1,N_2}^{(1,1)}(t)+\frac{1}{N_1+N_2}\right).
\end{equation}
The only remaining term, $(pp,qp$), is the most important: in this case, only one $q$ is surely not enough to re-create $\alp$ and thus, some cancellation is needed to close the Gr\"onwall estimate \eqref{eq:gronwall}. Indeed, the key fact is that
\[
p_1^BV_{12}(x_1-y_1)p_1^B=p_1^B\Big(V_{12}*|v|^2\Big)(x_1)p_1^B.
\]
This ``dressing'' of the true potential $V_{12}$ allows one to get an exact cancellation with the mean-field potential and obtain
\begin{equation}\label{eq:estimateI}
(pp,qp)+\text{c.c.}=0.
\end{equation}
Collecting \eqref{eq:estimateII}, \eqref{eq:estimateIII} and \eqref{eq:estimateI}, one finally gets \eqref{eq:gronwall}.

\section{Proof of Theorem \ref{theorem:mainGP}}

To describe how the proof proceeds, we need to revisit more in detail the so-called ``counting'' method developed by Pickl. In order to get more compact expressions, we drop the subscript $N_1,N_2$ in $\alpha$, $\Psi$, $\mathcal{E}$; the reader should keep in mind that everything always depends on the two population numbers. Given $p_1^A$ and $q_1^A$ as in \eqref{eq:def_p} and \eqref{eq:def_q}, we define a new family of projectors: for each $k\in N$, take
\begin{equation} \label{eq:defPk}
P_k^A:=\Big(q_1^A\dots q_k^Ap_{k+1}^A\dots p_{N}^A\Big)_{sym},
\end{equation}
with the convention that $P_k^A=0$ if $k>\mathbb{N}$ or $k<0$; we remark that the symbol `sym' in \eqref{eq:defPk} denotes the mere sum (without normalisation factor) of all possible permuted versions of the considered string of projections. A perfectly analogous definition of $P_k^B$ of course holds for the sector $B$.  By definition, the range of $P_k^A$ is the component of the Hilbert space in which exactly $k$ particles of type $A$ are in a state orthogonal to $u$ (recall that $p=|u\rangle\langle u|$), that is to say outside of the condensate. Thus, $\|\,P_k^A\;\Psi\,\|^2=\langle\,\Psi_{N_1,N_2},\;P_k\,\Psi_{N_1,N_2}\,\rangle$ is a measure of how large the component of $\Psi_{N_1,N_2}$ is, with exactly $k$ particles of type $A$ outside the condensate.

Now, given a positive function $g:\mathbb{N}\rightarrow\mathbb{R}$, define the operator
\begin{equation}
\widehat{g}^{\,A}:=\sum_{k=0}^{N_1}g(k)\,P_k^A,
\end{equation}
and the functional
\begin{equation}
\alpha_{N_1,N_2,g}^{(1,0)}:=\langle\,\Psi_{N_1,N_2},\;\widehat{g}^{\,A}\,\Psi_{N_1,N_2}\,\rangle.
\end{equation}
This amounts to assign some weight $g(k)$ to the component of a many-body state with exactly $k$ particles of type $A$ outside the condensate, and then summing over $k$. In the same way one defines
\begin{equation}
\alpha_{N_1,N_2,g}^{(0,1)}:=\langle\,\Psi_{N_1,N_2},\;\widehat{g}^{\,B}\,\Psi_{N_1,N_2}\,\rangle.
\end{equation}
The interest in this construction of course depends on the choice of $g$; it turns out that for some $g$'s, convergence to zero of both $\alpha_{N_1,N_2,g}^{(1,0)}$ and $\alpha_{N_1,N_2,g}^{(0,1)}$ is equivalent to convergence in trace norm \eqref{eq:tesiGP}. This is true, for example, for the special choice of the weight function $s(k):=k/N$, which yields to the single-component analogous of \eqref{eq:def_alpha}.

\subsection{The functional $\alpha_{m,<}^{(1,0)}$}
Unfortunately, the scaling in \eqref{eq:HN1N2GP} is too singular to allow one to close a Gr\"onwall argument for the weight $s(k)$. We try to explain here all the modifications needed in order to get the machinery working. It turns out that, if one tries to perform calculations with the weight $s$, one gets
\[
\big|\de_t\langle\,\Psi,\;\widehat{s}^{\,A}\;\Psi\,\rangle\big|\leqslant C\Big(\langle\,\Psi,\;\widehat{s}^{\,A}\;\Psi\,\rangle+\langle\,\Psi,\;\widehat{n}^{\,A}\;\Psi\,\rangle+o(1)+\big|\mathcal{E}(\Psi)-\mathcal{E}^{GP}(u,v)\big|\Big),
\]
where $n(k):=(k/N)^{1/2}$. Since $n(k)\geqslant s(k)$, the summand $\langle\,\Psi,\;\widehat{n}\;\Psi\,\rangle$ cannot be bounded and the estimate cannot be closed. This would suggest, in principle, that a Gr\"onwall estimate could be proven only by choosing as functional to control
\begin{equation}
\widetilde\alpha^{(1,0)}:=\langle\,\Psi,\;\widehat{n}^{\,A}\;\Psi\,\rangle+\big|\mathcal{E}(\Psi)-\mathcal{E}^{GP}(u,v)\big|.
\end{equation}
We observe that the convergence to zero of such $\widetilde{\alpha}$ would allow again to obtain the statement in trace norm \eqref{eq:tesiGP}, since (see Lemma 6.1 in \cite{p-external2015}, adaptable to the two-component case)
\[
\lim_{\substack{N_1\rightarrow\infty \\ N_2\rightarrow\infty}}\langle\,\Psi,\;\widehat{n}\;\Psi\,\rangle=0\quad \Leftrightarrow\quad \lim_{\substack{N_1\rightarrow\infty \\ N_2\rightarrow\infty}} \gamma^{(1,1)}(t)=\;|u_t\otimes v_t\rangle\langle u_t\otimes v_t|.
\]
The functional $\widetilde{\alpha}$ is however not efficient enough yet; the reason is that its first and second derivative, which crucially enter in computations (see again \cite{p-external2015}, Appendix A.2), are singular for $k=0$. For this reason, one defines a new weight, with a less singular behavior for small $k$'s. For some fixed $\xi>0$, we define
\begin{equation}
m(k):=\begin{cases}
\sqrt{k/N},\qquad\qquad\qquad\quad\text{for } k\geqslant N^{1-2\xi}\\
$\,$\\
\frac{1}{2}\big(N^{-1+\xi}k+x^{-\xi}\big),\qquad\, \text{ else}.
\end{cases}
\end{equation}
With this weight, we define a new functional as
\begin{equation}\label{eq:def_alphaGP<}
\alpha_{m,<}^{(1,0)}:=\langle\,\Psi,\;\widehat{m}^{\,A}\;\Psi\,\rangle+\big|\mathcal{E}(\Psi)-\mathcal{E}^{GP}(u,v)\big|.
\end{equation}
The vanishing of this indicator and of its corresponding $\alpha_{m,<}^{(0,1)}$ is again equivalent to convergence in trace norm since
\[
n(k)\leqslant m(k)\leqslant\text{max}\{n(k),N^{-\xi}\}.
\]
It turns out that $\alpha_{m,<}^{(1,0)}$ and $\alpha_{m,<}^{(0,1)}$ allow to control convergence for the softer scaling 
\begin{equation}
V_N= N^{-1+3\beta}V(N^\beta(x-y)),
\end{equation}
with $0<\beta<1$, but \emph{not} for the true Gross-Pitaevskii scaling, corresponding to the case $\beta=1$. The reason is that, for $\beta=1$, an important role is played by the short-scale correlation among particles.

\subsection{Adding correlations}

In the derivation of Gross-Pitaevskii equation, correlations are customarily accounted for (see for example \cite{bdos-quant2012}) by means of the solution $f_N$ to the zero-energy scattering equation
\begin{equation}\label{eq:zero_energy_true}
\left(-\Delta_x+\frac{1}{2}V_N(x)\right)f_N(x)=0, \qquad\text{with }f(x)\rightarrow 1\,\,\,\text{for }|x|\rightarrow\infty,
\end{equation}
where $V_N(x)=N^2V(Nx)$. In the setting we are considering, it is however more efficient~\cite{p-external2015} to consider a slight modification of \eqref{eq:zero_energy_true}. Recalling that we defined $a_k$ as the scattering lenght of $V_k$ for $k\in\{1,2,12\}$, we can define, for given constants $C_j, C_{12}$, the new potentials
\begin{equation}
W_{j,\beta}(x):=\begin{cases}
\dfrac{4\pi a_j}{N_j}N_j^{3\beta},\qquad\text{for } N_j^{-\beta}<x<C_jN_j^{-\beta}\\
\\
\;0 \qquad\qquad\quad\,\;\;\text{else},
\end{cases}
\end{equation}
with $j\in\{1,2\}$, and
\begin{equation}\hspace{-0.03cm}
W_{12,\beta}(x):=\begin{cases}
4\pi a_{12}(N_1+N_2)^{3\beta-1},\,\,\;\;\, (N_1+N_2)^{-\beta}<x<C_{12}(N_1+N_2)^{-\beta}\\
\\
\;0 \qquad\qquad\qquad\qquad\quad\,\text{else}.
\end{cases}
\end{equation}
One can show that there exist $C_j$, $C_{12}$ such that the scattering lengths of $N_j^2V(N_j^\beta\cdot)-W_{j,\beta}(\cdot)$ and of $(N_1+N_2)^2V((N_1+N_2)^\beta\cdot)-W_{12,\beta}(\cdot)$ are zero (see Lemma 5.1 in \cite{p-external2015} or Lemma 5.5 in \cite{jlp-two2016} for a more detailed proof). One can now define two functions $f_{j,\beta}$ and $g_{j,\beta}$, $j=1,2$, by means of a modified zero-energy scattering equation, namely
\begin{equation}\label{eq:zero_energy_mod}
\left(-\Delta_x+\frac{1}{2}\big(V_{j,N_j}(x)-W_{j,\beta}(x)\big)\right)f_{j,\beta}(x)=0, \qquad\text{with }f_{j,\beta}(x)\rightarrow 1\,\,\,\text{for }|x|\rightarrow\infty,
\end{equation}
and
\begin{equation}\label{eq:def_g}
g_{j,\beta}:=1-f_{j,\beta},
\end{equation}
with the analogous definition for $f_{12,\beta}$ and $g_{12,\beta}$. By insertion of the new potential, it turns that out the norms of $g_{j,\beta}$ have a better behavior in $(N_1+N_2)$ than they would have without the additional potential.

Now, by construction, the key properties of $W_{j,\beta}$ are
\[
\| W_{j,\beta}\|_1\sim O(N_1+N_2)^{-1},\quad\text{and}\quad\|W_{j,\beta}\|_\infty\sim O(N_1+N_2)^{-1+3\beta},
\]
and the same holds for $W_{12,\beta}$. For this reason, replacing $V_{j_N,j}$ (respectively $V_{12,N_1+N_2}$) in the proof with $W_{j,\beta}$ (respectively $W_{12,\beta}$) one would deal with a potential with a much less peaked scaling; of course the price to pay is the appearance of their difference, but this can be dealt with by adding a further term to the functional one aims to control.
\begin{definition}[$\alpha_m^{(1,0)}$ and $\alpha_m^{(0,1)}$] \label{def:indicators}
	We define \emph{the} indicators of convergence for the Hamiltonian \eqref{eq:HN1N2GP} as
	\begin{equation}\label{eq:def_alphaGP1}
	\begin{split}
	\alpha_m^{(1,0)}:=\alpha_{m,<}^{(1,0)}&-N_1(N_1-1)\text{Re}\langle\,\Psi,\;g_{1,\beta}(x_1-x_2)\;R^{\,A}_{(12)}\psi\,\rangle\\
	&-N_1N_2\text{Re}\langle\,\Psi,\;g_{12,\beta}(x_1-y_1)\;R^{\,A}_{(12)}\psi\,\rangle
	\end{split}
	\end{equation}
	and
	\begin{equation}\label{eq:def_alphaGP2}
	\begin{split}
	\alpha_m^{(0,1)}:=\alpha_{m,<}^{(0,1)}&-N_2(N_2-1)\text{Re}\langle\,\Psi,\;g_{2,\beta}(y_1-y_2)\;R^{\,B}_{(12)}\psi\,\rangle\\
	&-N_1N_2\text{Re}\langle\,\Psi,\;g_{12,\beta}(x_1-y_1)\;R^{\,B}_{(12)}\psi\,\rangle,
	\end{split}
	\end{equation}	
	where $R_{(12)}:=p_1p_2(\widehat{m}-\widehat{m}_2)+(p_1q_2+q_1p_2)(\widehat{m}-\widehat{m}_1)$, having used the shorthand notation $\widehat{m}_j:=\sum_{k=0}^Nm(k)P_{k+j}$.
\end{definition}

\begin{remark}
	The terms $\widehat{m}-\widehat{m}_1$ and $\widehat{m}-\widehat{m}_2$ are bounded in operator norm by $\sup_k|m'(k)|$. This is the reason why we had to define $m(k)$ by cutting $(k/N)^{1/2}$ for small $k$'s.
\end{remark}

\begin{remark}
	The terms subtracted from $\alpha_{m,<}^{(1,0)}$ and $\alpha_{m,<}^{(0,1)}$ in Def. \ref{def:indicators} are real but with no definite sign. However, one can easily prove a priori estimates for them; for example
	\begin{equation}\label{eq:apriori}
	N_1(N_1-1)\text{Re}\langle\,\Psi,\;g_{1,\beta}(x_1-x_2)\;R^{\,A}_{(12)}\psi\,\rangle\le N^{-\eta},
	\end{equation}
	for some $\eta>0$, and the same holds for the other four terms. This helps in closing the Gr\"onwall estimate even though the considered functionals have no definite sign.
\end{remark}

By repeating the computations in Appendix A.2 in \cite{p-external2015} with minor changes, one can prove the estimate
\[
\frac{d}{dt}\Big(\alpha_m^{(0,1)}(t)+\alpha_m^{(1,0)}(t)\Big)\leqslant f(t)\Big(\alpha_{m,<}^{(1,0)}(t)+\alpha_{m,<}^{(0,1)}(t)+(N_1+N_2)^{-\eta}\Big).
\]
Now, by using the a priori estimate \eqref{eq:apriori} and Gr\"onwall Lemma, this is enough to get
\[
\alpha_{m,<}^{(1,0)}(t)+\alpha_{m,<}^{(0,1)}(t)\leqslant e^{\int_0^tf(s)ds}\Big(\alpha_m^{(0,1)}(0)+\alpha_m^{(1,0)}(0)+N^{-\eta}\Big).
\]
Since $\alpha_m^{(0,1)}(0)+\alpha_m^{(1,0)}(0)$ is converging to zero by Assumption (B3) and by Eq. \eqref{eq:chain}) for $t=0$, we get the thesis by using again Eq. \eqref{eq:chain} for $t>0$.

\section{Spinor condensates and other multi-component models}\label{sect:open}

As already remarked, the study of multi-component condensates is a very popular topic in theoretical and experimental physics; we would like to present in this Section an account of some highly studied models, different from the mixture gas considered in this paper, that fall under the name of multi-component condensates.

In Sect. \ref{sect:multi} we mentioned that a well-known example of multi-component condensate is a gas of spin bosons. Consider for example a system of atoms allowed to populate different hyperfine states; it is often assumed (and easily realizable with modern experimental techniques), that an external field is tuned in such a way that only two hyperfine levels are coupled and enter the effective Hamiltonian. When this is the case, then one can model the system by means of an auxiliary spin-1/2 bosonic theory. These systems are often referred to as \textit{pseudo-spinor condensates}, since a proper spin-spin interaction is not present; nonetheless, the situation is already non trivial since one could even account for transitions between the two hyperfine levels: this can be realized for example by a (possibly time-dependent) external magnetic field. In this setting, the effective equations for the spin-1/2 case are (see for example \cite[Sect. 21.3]{ps2016})
\begin{equation}\label{eq:rabi} 
\begin{split}
&i\de_t u_t=-\Delta u_t+8\pi a(|u_t|^2+|v_t|^2)u_t+B(t)v_t\\
&i\de_t v_t=-\Delta v_t+8\pi a(|u_t|^2+|v_t|^2)v_t+B(t)u_t,
\end{split}
\end{equation}
where $a$ is the scattering length of the interaction and $B(t)$ is the magnetic field; the linear coupling provided by $B(t)$ is called \emph{Rabi coupling}. We refer the reader to \cite{mo-ps2017} for the derivation of \eqref{eq:rabi} from the many-body dynamics of a pseudo-spinor condensate.

An even more interesting situation is the presence of spin-spin interaction. In the relevant case of a gas of alkali atoms, one should in principle take into account the presence of different values of hyperfine spin (e.g. $F=1$ and $F=2$); however, due to energetic arguments, a good low-energy approximation for the interaction can be obtained by completely neglecting the presence of one of the two hyperfine level, say $F=2$. Under this approximation, it turns out that a general interaction Hamiltonian that preserves the hyperfine spin of the individual atoms and is rotationally invariant in the hyperfine spin space has the form
\begin{equation}
\delta (x_i-x_j)\big(c_0+c_1\mathbf{S_i}\cdot\mathbf{S_j}\big),
\end{equation}
where $\mathbf{S_i}$ is the vector of spin-1 operators for the particle $i$. This not only provides population transfer, but it also correlates particles and for this reason the effect must be present on the non-linearity too. The factor $\delta(x_i-x_j)$ can be modeled by some Gross-Pitaevskii potential with scattering length $c_1$, and thus we can write the total spin-spin interaction term for a \emph{spinor condensate} (neglecting the irrelevant $c_0$) as
\begin{equation}
N^2\sum_{i<j}V(N(x_i-x_j))\mathbf{S_i}\cdot\mathbf{S_j}.
\end{equation}
For the spin-1 case, this produces the equations \cite{ho-prl, om-jpsj}
\begin{equation}\label{eq:spin_spin}
\begin{split}
&i\de_t u_t=-\Delta u_t+8\pi a\Big(|v_t|^2u_t+\overline{w}_t v_t^2+|u_t|^2u_t-|w_t|^2u_t\Big)\\
&i\de_t v_t=-\Delta v_t+8\pi a\Big(|u_t|^2v_t+2\overline{v}_tw_tu_t+|w_t|^2v_t\Big)\\
&i\de_t w_t=-\Delta w_t+8\pi a\Big(|v_t|^2w_t+\overline{u}_t v_t^2-|u_t|^2w_t+|w_t|^2w_t\Big),
\end{split}
\end{equation}
where again $a$ is the scattering lenght of $V$. The rigorous derivation of the system  \eqref{eq:spin_spin} from many-body quantum dynamics is undoubtedly one of the next frontiers in the mathematics of the Bose gas.

\begin{acknowledgement}

{Partially supported by the 2014-2017 MIUR-FIR grant ``\emph{Cond-Math: Condensed Matter and Mathematical Physics}'', code RBFR13WAET and by Gruppo Nazionale per la Fisica Matematica (GNFM-INdAM).}
\end{acknowledgement}
%

\end{document}